# Few-layer Nanoplates of $Bi_2Se_3$ and $Bi_2Te_3$ with Highly Tunable Chemical Potential

Desheng Kong[1], Wenhui Dang[2], Judy J. Cha[1], Hui Li[2], Stefan Meister[1], Hailin Peng[2,*], Zhongfan Liu[2], Yi Cui[1,*]

[1]Department of Materials Science and Engineering, Stanford University, Stanford, CA 94305, USA.
[2]Center for Nanochemistry, Beijing National Laboratory for Molecular Sciences (BNLMS), State Key Laboratory for Structural Chemistry of Unstable and Stable Species, College of Chemistry and Molecular Engineering, Peking University, Beijing 100871, P. R. China.

*To whom correspondence should be addressed. E-mail: H.P., hlpeng@pku.edu.cn, and Y.C., yicui@stanford.edu.

Topological insulator (TI) represents an unconventional quantum phase of matter with insulating bulk bandgap and metallic surface states. Recent theoretical calculations and photoemission spectroscopy measurements show that Group V-VI materials $Bi_2Se_3$, $Bi_2Te_3$ and $Sb_2Te_3$ are TI with a single Dirac cone on the surface. These materials have anisotropic, layered structures, in which five atomic layers are covalently bonded to form a quintuple layer, and quintuple layers interact weakly through van der Waals interaction to form the crystal. A few quintuple layers of these materials are predicted to exhibit interesting surface properties. Different from our previous nanoribbon study, here we report the synthesis and characterizations of ultrathin $Bi_2Te_3$ and $Bi_2Se_3$ nanoplates with thickness down to 3 nm (3 quintuple layers), via catalyst-free vapor-solid (VS) growth mechanism. Optical images reveal thickness-dependant color and contrast for nanoplates grown on oxidized silicon (300nm $SiO_2$/Si). As a new member of TI nanomaterials, ultrathin TI nanoplates have an extremely large surface-to-volume ratio and can be electrically gated more effectively than the bulk form, potentially enhancing surface states effects in transport measurements. Low temperature transport measurements of a single nanoplate device, with a high-k dielectric top gate, show decrease in carrier concentration by several times and large tuning of chemical potential.

The growth and physical properties of $A_2B_3$ (A=Bi, Sb; B=Se, Te) chalcogenide materials have been studied for more than half a century due to their interesting thermoelectric properties.[1-3] Recently, however, research interest on this family of compounds increased dramatically because $Bi_2Se_3$, $Bi_2Te_3$ and $Sb_2Te_3$ were discovered to be three-dimensional (3D) topological insulators (TIs), a new quantum matter with conductive massless Dirac Fermions on their surface.[4] In TI, strong spin-orbit coupling induces non-trivial surface states that are robustly protected against any time reversal perturbations, such as crystal defects and non-magnetic impurities.[5-9] The distinct properties of TI surface states are attractive for fundamental research as well as applications in spintronics, quantum information and low energy dissipation electronics. Experimentally, angle resolved photoemission spectroscopy (ARPES) measurements on bulk single crystals have verified the existence of the 3D TI and spin-momentum locked Dirac Fermion nature of the surface states.[10-13]

$Bi_2Se_3$, $Bi_2Te_3$ and $Sb_2Te_3$ share the same layered rhombohedral crystal structure in space group $D^5_{3d}$ (R3m). Each charge neutralized layer is formed by five covalently bonded atomic sheets, for example Se-Bi-Se-Bi-Se in $Bi_2Se_3$, defined as a quintuple layer (QL) with thickness of ~1nm (Figure 1A). The quintuple layers are weakly bonded together by van der Waals interaction to form the crystal. These anisotropic crystallography characteristics are significant for controllable growth of TI materials.

Synthesizing low-dimensional structures out of this family of compounds is attractive for TI study. Compared with bulk samples, nanostructured TI materials are a compelling system for probing TI surface states due to their large surface-to-volume ratio, in favor of manifesting the surface states during transport measurements. For example, Aharonov-Bohm oscillation was recently observed in $Bi_2Se_3$ nanoribbons by some of the present authors, providing direct transport evidence of the robust, conducting surface states.[14, 15] Kondo effect on magnetically doped $Bi_2Se_3$ nanoribbons has also been observed.[16] In addition, theoretical calculations and ARPES measurements on ultrathin molecular beam epitaxy (MBE) films of $Bi_2Se_3$ and $Bi_2Te_3$ reveal the unconventional size effects: evolution from 2D TI to 3D TI phase depending on film thicknesses.[17, 18] Thin film form of TI can be obtained by MBE growth,[18, 19] but MBE is expensive and less accessible. Mechanical exfoliation of thin sheets from bulk crystals,[20] widely used for obtaining graphene from graphite, can also achieve thin TI layers, but with very low yield and irregular shapes. Nanoribbons and nanowires of chalcogenide materials are easily synthesized by vapor-liquid-solid (VLS) process,[14, 16, 21-24] but only with moderate thickness range (~30-100 nm). Facile synthesis method to obtain ultrathin TI materials down to few QLs is desired. Here, we developed a catalyst-free vapor-solid (VS) synthesis process to grow ultrathin nanoplates (NPs) of $Bi_2Se_3$ and $Bi_2Te_3$ with thicknesses down to ~3nm (3QLs). These NPs show a strong electrical gating response and highly tunable chemical potential in transport measurements, attractive for probing TI surface states and future device applications.

Synthesis of $Bi_2Se_3$ and $Bi_2Te_3$ NPs is carried out in a 12 inch horizontal tube furnace (Lindberg/Blue M) via catalyst-free vapor transport and deposition process (Figure 1B), using a set-up similar to our previous nanoribbon synthesis by VLS growth.[14, 16, 24-26] Powder of source materials (99.999% $Bi_2Se_3$ or $Bi_2Te_3$) is placed in the hot center of the tube furnace. Oxidized silicon wafers (300nm $SiO_2$/Si) are placed at the cold zone (~12 cm away from the center) as growth substrates. For the synthesis, the tube is initially pumped down to a base pressure of about 100 mTorr, and flushed with ultrapure Ar several times to remove oxygen residue. With Ar gas flow (50 sccm for $Bi_2Se_3$ and 20 sccm for $Bi_2Te_3$), the hot center of the furnace is heated to growth temperature (460~500°C for $Bi_2Se_3$ and 450~480 °C for $Bi_2Te_3$) and is held at the temperature and a certain pressure (100 torr for $Bi_2Se_3$ and 20 torr for $Bi_2Te_3$) for about 5 min, followed by natural cooling down. The $SiO_2$/Si substrate is in the temperature range of 320 °C ~360 °C.

The synthesized products are characterized by scanning electron microscopy (SEM, FEI XL30 Sirion and Hitachi S-4800) and atomic force microscopy (AFM, Park Systems XE-70 and Digital Instruments Nanoscope IIIA). Figure 2 shows a SEM image of $Bi_2Te_3$ NPs (SEM image of $Bi_2Se_3$ NPs is provided in Supporting Information, Figure S1). NPs usually exhibit triangular or hexagonal morphologies with lateral dimensions extending from several micrometers up to ~20 μm. On average, the lateral dimensions of $Bi_2Te_3$ NPs are larger than those of $Bi_2Se_3$ NPs. We observe 60° or 120° facets at the edges, as expected from their crystal structure. SEM images show that these NPs grow directly on oxidized Si surface via VS growth mechanism, without catalyst particles at the ends of NPs. They are strongly bonded to the substrate and hardly removed during sonication in organic solvents. The thickness of synthesized NPs is measured by AFM, shown in Figure 3A as an example. Most of the NPs have very flat surfaces with a uniform thickness across the lateral dimensions (Figure 3B for $Bi_2Te_3$ and 3C for $Bi_2Se_3$, respectively). The thickness distribution for $Bi_2Te_3$ is obtained by scanning a substrate area of ~200 μm$^2$ (Figure 3D). NPs with uniform thicknesses between 3QLs to 6QLs (3-6 nm) are major products. We also observe minority NP products whose thickness is not uniform but with 1–4 QL steps (Supporting Information, Figure S2, and Figure 4B, 4C).

Contrast difference in optical images provides a facile way to identify ultrathin TI NPs of different thicknesses. Optical images are acquired by Leitz ERGOPLAN microscope (10X eyepieces combined with 50X and 150X objective lens), under white light illumination conditions. As-grown ultrathin TI NPs are readily observable on oxidized silicon substrates (300nm $SiO_2$/Si), as shown in Figure 4A (50X objective lens). Here, we establish the relation between optical contrast and the thickness of TI NPs, by correlating optical images (Figure 4B, with 150X objective lens) with AFM data (Figure 4C) for a NP with multiple single QL steps. As thickness increases from 4QL to 13QL (marked by black numbers in Figure 4C), the optical colors of the NP in Figure 4B gradually change from dark purple to yellow. For TI NPs less than 13QL, the optical contrast variation allows differentiating thickness difference in a step of 1QL. Similar to identifying few-layer graphene deposited on oxidized silicon,[27] $SiO_2$ thickness is critical to enhance the optical contrast difference of ultrathin NPs of various thicknesses. Physical origin of the large optical contrast is ascribed to strong amplitude modulation of reflection at the air-2D TI crystal-$SiO_2$ interface.[28]

The reason that the VS growth mechanism can produce ultrathin NPs (3-6nm) is due to the anisotropic bonding nature of $Bi_2Se_3$ and $Bi_2Te_3$. Nucleation takes place randomly on the $SiO_2$ surface and more probably on physical or chemical inhomogeneous sites of substrate, such as surface roughness and impurities. Nuclei have the c crystal axis perpendicular to the substrate surface. After nucleation, the growth process continues with a high anisotropic nature (Figure 1C). The atoms at the edge of the nuclei have dangling bonds ready to bind covalently with incoming atoms. The top surface is terminated with chemically saturated Se atoms. Atoms adsorbed onto the top surface from gas phase cannot form covalent bonds with the top Se atoms, and consequently tend to diffuse around to find the edges. Therefore, the lateral dimension grows much faster than the vertical thickness dimension. The high growth anisotropy results from the bonding anisotropy. Using the ratio of the lateral to vertical dimension of NPs, we estimate that the growth anisotropy is several thousand. In fact, nucleation of additional quintuple layers on top of the existing NPs does take place slowly. Additional nucleations are believed to happen close to the edge of existing NPs, since we were able to catch some NPs with the edge thicker than the center (Figure 4B, 4C and Supporting Information, Figure S2).

To characterize the crystal structure and chemical compositions of synthesized products, NPs are studied by a 200 kV FEI Tecnai F20 transmission electron microscope (TEM) and energy-dispersive X-ray spectroscopy (EDX) equipped inside the TEM. TEM sample preparation details are provided in Supporting Information. Figure 5A shows a typical low magnification TEM image of $Bi_2Te_3$ NPs grown on 100nm $SiN_x$ membrane. Corresponding high resolution TEM (HRTEM) lattice fringes and spot pattern of selected area electron diffraction (SAED), shown in Figure 5B and 5C, demonstrate the single crystalline nature of the NPs. HRTEM image reveals expected hexagonal lattice fringes with a lattice spacing of 0.22 nm, consistent with the lattice spacing of (11-20) planes. Similar to VLS nanoribbons, NPs have (0001) facets as top and bottom surfaces, and (01-10) facets as side surfaces.[14, 16, 24] The TEM image of $Bi_2Se_3$ NPs transferred to a lacey carbon support film on a TEM copper grid is shown in Figure 5D. The HRTEM image and corresponding FFT pattern, shown in Figure 5E and 5F, prove it is also single crystal with similar structural characteristics to $Bi_2Te_3$ NPs. EDX analysis out of single NPs confirms the correct stoichiometry, within the accuracy of the measurement (Supporting Information, Figure S3).

Electrical transport properties are studied in single NP devices with a top gate. Here, we use $Bi_2Te_3$ NP devices to demonstrate the tunable chemical potentials and pronounced weak antilocalization effect. Ohmic contacts to NPs are fabricated by standard e-beam lithography and thermal evaporation of Cr/Au contacts (10nm/160nm). Thin high-k dielectric layer (~30nm) of $Al_2O_3$ is deposited by atomic layer deposition (Cambridge NanoTech Inc., details in Supporting Information). E-beam lithographically patterned Cr/Au (10nm/160nm) metal top gate is made by thermal evaporation afterwards. An optical image of a typical device is shown as the right inset of Figure 6A. Transport measurements are performed at 2K in Quantum Design PPMS-7 instrument, combined with a digital lock-in amplifier (Stanford Research Systems) and a Keithley 2400 source meter (set-up details are provided in Supporting Information). For resistance measurement, a four-point configuration is used to remove contribution from the contact resistance. The change in four-point sheet resistance $R_S$ with top gate voltage $V_G$ is shown in Figure 6A. The gating response indicates the NPs are a n-type doped semiconductor, similar to the case of $Bi_2Se_3$ nanoribbons.[14, 16, 24] The n-type nature is usually ascribed to Te/Se vacancies and anti-site defects formed during growth process.[11, 13] For a typical NP device, the resistance (Figure 6A) increases quasi-linearly with negative gate voltage up to -17.5V, beyond which the resistance rises more rapidly when the gate voltage further increases towards negative direction. The inset of Figure 6A shows the slope of the resistance as a function of the gate voltage, demonstrating the sharper increase of the resistance beyond -17.5V. This enhancement of the gating effects strongly suggests that the chemical potential is approaching the bottom of conduction band at the gate voltage of -17.5V, beyond which a large portion of bulk electrons are being depleted by additional negative gate voltage. The Hall measurements (Figure 6B) show the areal carrier density of the NP is $7.2 \times 10^{13}$ cm$^{-2}$ at zero gate voltage, and $2.1 \times 10^{13}$ cm$^{-2}$ at -20V gate voltage (change by a factor of 3). Hall measurements confirm as-grown NPs are a heavily doped n-type semiconductor. Nevertheless, NP devices show much stronger gate response than previously reported nanoribbons and mechanical exfoliated samples.[29, 30] The pronounced gate effects and tunable chemical potential are achieved by the combination of a strong electrical field from the top gate using a high-k dielectric layer and ultrathin thickness of NPs from VS growth. The magnetoconductance of the same NP device is shown in Figure 6C, where the applied magnetic field (B) is perpendicular to the NP plane. At low magnetic field |B|<0.80T, the weak antilocalization effect manifests as a sharp peak in the magnetoconductance trace, resulting from strong spin orbit coupling in TI materials.[14, 31] The oscillations in conductance (Inset of Figure 6C) show aperiodic field dependence, as universal conductance fluctuation (UCF) commonly observed in mesoscopic metallic and semiconducting structures.[32] It is also observed in our previous $Bi_2Se_3$ nanoribbon measurement.[14] We expect that further lowering of the bulk carrier concentration may reveal interesting physics from TI surface states.

In summary, we develop a VS synthesis process to grow few-layer nanoplates of topological materials $Bi_2Se_3$ and $Bi_2Se_3$, with thicknesses down to ~3nm, or 3 quintuple layers. Optical images reveal thickness-dependant color and contrast for NPs grown on top of oxidized silicon surface (300nm $SiO_2$/ Si), similar to the case of graphene. Single NP devices with high-k dielectric top gate configuration show a highly tunable chemical potential and a pronounced weak antilocalization effect.

**Acknowledgement**. Y. C. acknowledges the support from the Keck Foundation. This work is also made possible by the King Abdullah University of Science and Technology (KAUST) Investigator Award (No. KUS-l1-001-12). H. P. acknowledges the support from NSFC (20973007, 20973013, 50821061) and MOST (2007CB936203).

**Supporting Information Available:** Additional SEM image, AFM image, EDX spectra, TEM sample preparations, ALD $Al_2O_3$ film growth conditions and transport measurement equipments set-up. This material is available free of charge via the Internet at http://pubs.acs.org.

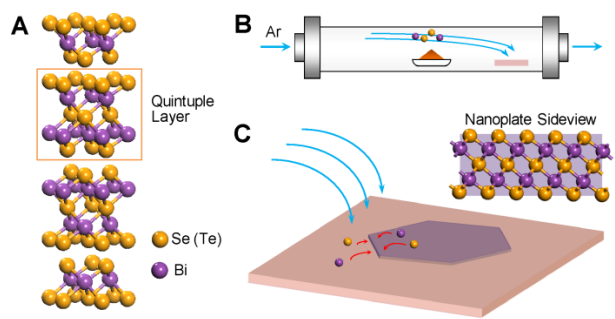

Figure 1. (A) Layered crystal structure of $Bi_2Se_3$ and $Bi_2Te_3$, with each quintuple layer (QL) formed by five Bi and Se (or Te) atomic sheets. (B) A schematic drawing of vapor-solid (VS) growth process for few-layer NPs of $Bi_2Se_3$ and $Bi_2Te_3$ in a horizontal tube furnace. (C) A schematic drawing of NPs growth mechanism, involving gas phase atoms diffused and attached to the side surface. Sideview of the NPs (1QL) show the Se terminated top and bottom surfaces with saturated bonds, and side surfaces with dangling bonds ready to bind with incoming atoms.

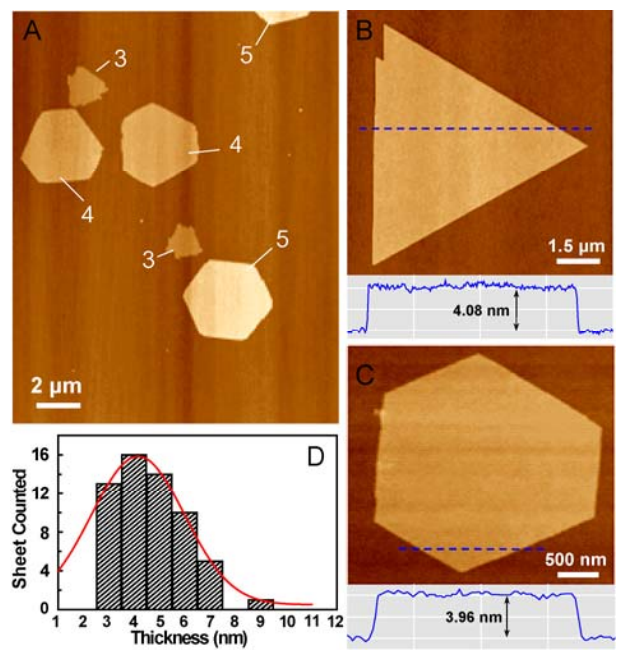

Figure 3. AFM characterizations of a few QL $Bi_2Te_3$ and $Bi_2Se_3$ NPs. (A) Overview of several ultrathin $Bi_2Te_3$ NPs. The numbers indicate the number of quintuple layers. (B) AFM image and a height profile (corresponding to the dashed line in the image) of a $Bi_2Te_3$ NP with 4QL. (C) AFM image and a height profile (corresponding to the dashed line in the image) of a $Bi_2Se_3$ NP with 4QL. (D) Histogram of $Bi_2Te_3$ NP thickness obtained from substrate area of ~200 $\mu m^2$. The smooth curve is a Gaussian fit to the thickness distribution.

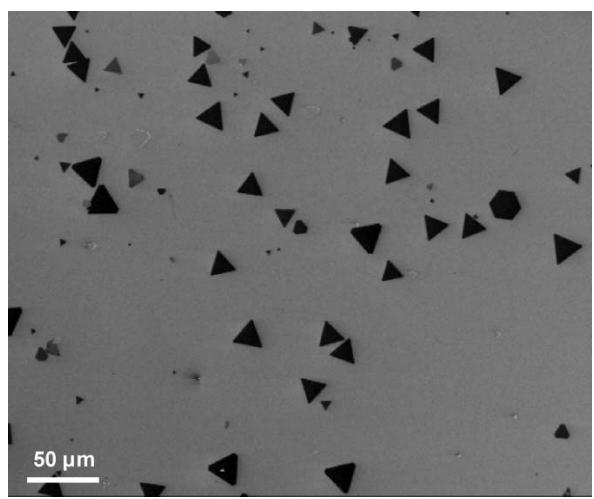

Figure 2. SEM image of $Bi_2Te_3$ NPs grown on oxidized silicon substrate (300nm $SiO_2$/Si). A SEM image of $Bi_2Se_3$ NPs is provided in Support Information (Figure S1).

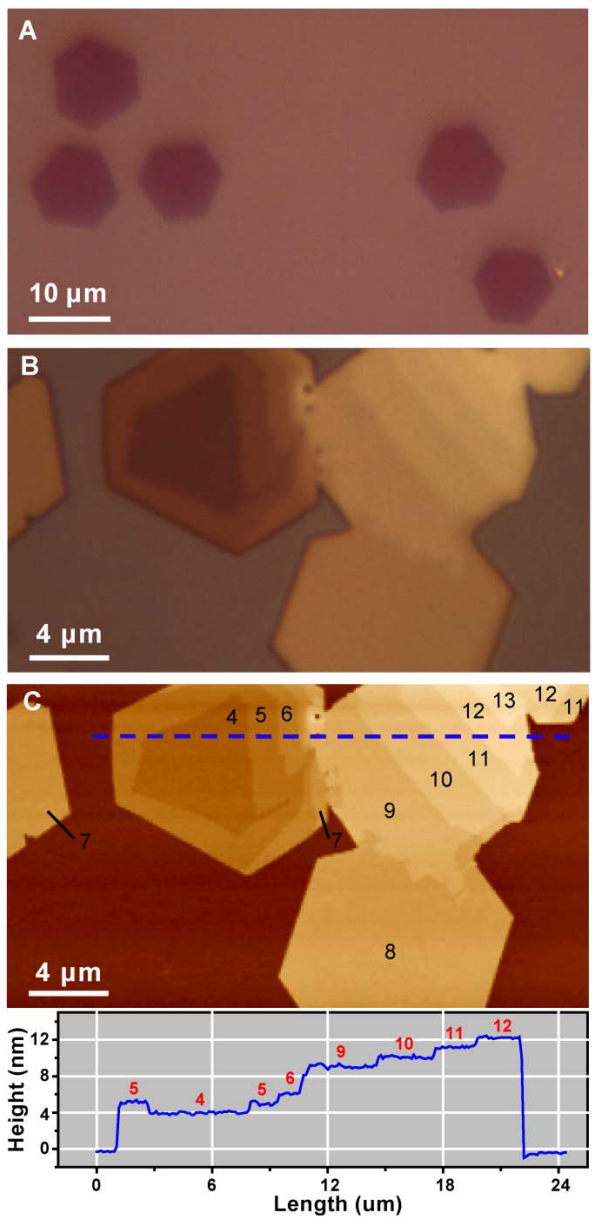

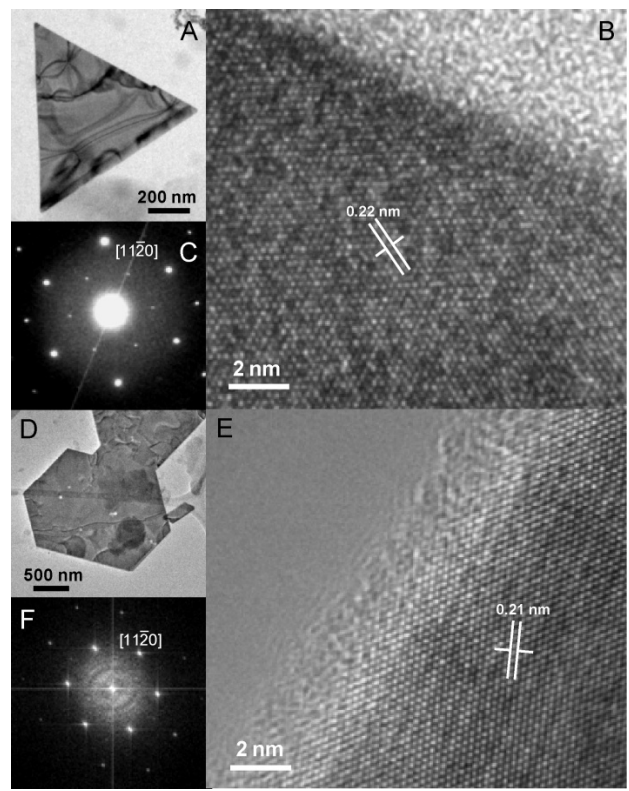

Figure 4. (A) A typical optical image of ultrathin NPs of $Bi_2Te_3$ grown on oxidized Si substrate (300nm $SiO_2$/Si) by 50X objective lens. NPs have clear optical contrast from the substrate for easy identification. (B) High magnification optical image (150X objective) of NPs with multiple step edges. The color changes from dark purple to yellow color, as the thickness increases from 4QL to 13QL. (C) An AFM image and height profile (corresponding to the dashed line-cut) from the same NPs shown in (B). The number of QLs is marked in AFM (black) and height profile (red). These NPs have multiple single QL steps, confirming layer-by-layer growth modes. By comparing between (B) and (C), optical images indeed provide sufficient contrast for differentiating each QL from 4QL to 14QL.

Figure 5. TEM characterizations of $Bi_2Te_3$ and $Bi_2Se_3$ NPs. (A) TEM image of a $Bi_2Te_3$ NP grown on 100nm $SiN_x$ membrane. HRTEM image (B) and SAED pattern (C) acquired from the same $Bi_2Te_3$ NP. (D) TEM image of a $Bi_2Se_3$ NP transferred to a lacey carbon support film on a TEM copper grid. (E) HRTEM image obtained from the same $Bi_2Se_3$ NP, with corresponding fast Fourier Transformation (FFT) pattern shown in (F).

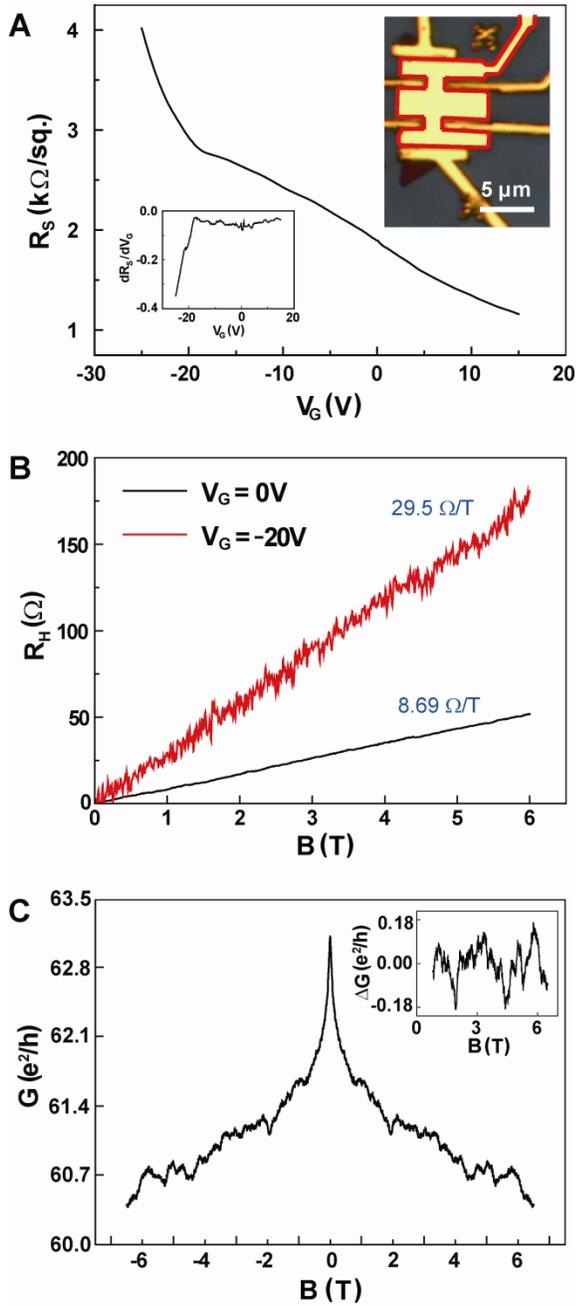

Figure 6. Electron transport measurement at 2K. (A) Device sheet resistance $R_S$ vs. top gate voltage $V_G$. $R_S$ inceases by ~210% at negative gate voltage of -25V. Differential resistance over gate voltage is shown in the inset on the left to highlight the enhancement of gating response for $V_G$ <-17.5V. A typical optical image of a NP device is shown in the inset on the right, in which the top gate is encircled by red line to guide the eye. (B) Hall measurements at $V_G$= 0V and $V_G$ = -20V. Areal carrier concentration decreases by a factor of 3 by applying -20V negative gate voltage. (C) Magnetoconductance in the field range of ±6.5 T. Conductance oscillations (Inset) is obtained by subtracting the background.